# Combining Design Thinking and Software Requirements Engineering to create Human-centered Software-intensive Systems

**Jennifer Hehn[1], Daniel Mendez[2, 3]**

1 Institute of Technology Management, Berne University of Applied Sciences

2 Blekinge Institute of Technology, Sweden

3 fortiss GmbH, Germany

**Abstract**  Effective Requirements Engineering is a crucial activity in software-intensive development projects. The human-centric working mode of Design Thinking is considered a powerful way to complement such activities when designing innovative systems. Research has already made great strides to illustrate the benefits of using Design Thinking for Requirements Engineering. However, it has remained mostly unclear how to actually realize a combination of both. In this chapter, we contribute an artifact-based model that integrates Design Thinking and Requirements Engineering for innovative software-intensive systems. Drawing from our research and project experiences, we suggest three strategies for tailoring and integrating Design Thinking and Requirements Engineering with complementary synergies.

## 1. Introduction

The success of any software-intensive system anchors in the question how well it reflects its users' needs (Maguire and Bevan 2002) and surrounding constraints. "Getting the requirements right" – which is often associated with term *Requirements Engineering* – is consequently seen as one of the most significant endeavours in development projects (Broy 2006; Robertson and Robertson 2013). It is typically associated with initial phases of a software development life cycle and its major aim is to decide upon the relevant functional and non-functional properties of software-intensive products. Elementary tasks towards this goal include the elicitation of requirements, their analysis and negotiation also in terms of reaching consensus among all relevant stakeholders, their specification to accommodate subsequent engineering activities, and their validation in terms of ensuring the requirements' quality (e.g. correctness and consistency, among other attributes).

In accordance with the terminology introduced by the International Requirements Engineering Board[1], Requirements Engineering, therefore, denotes the "systematic and disciplined approach to the specification and management of requirements with the goal of understanding stakeholders' desires and needs and minimising the risk of delivering a system that does not meet these desires and needs". Given the human-centric nature of software – in the end, software is made by humans for humans – Requirements Engineering is undoubtedly a critical determinant for software quality, regardless of how Requirements Engineering exactly manifests itself in practical settings[2]. In a world pervaded by software and where most of our daily routines are supported – if not dominated – by software-intensive systems, excellence in RE is a de-facto key. At the same time, many companies struggle with capturing the users' needs effectively, often leading to software-intensive systems which (1) either miss important requirements, (2) reflect incorrect requirements (or incorrect assumptions), or (3) which reflect - technically speaking - the correct functionality, but are still rendered unusable as they lack important non-functional properties from the perspective of their end users. This gives rise to the need for new approaches that allow for a more human-centred Requirements Engineering.

In the following, we first elaborate on difficulties and limitations of contemporary Requirements Engineering principles and approaches, before motivating their symbiotic relationship with Design Thinking to create software-intensive systems in such human-centred way. Exploring the relationship of both historically grown worlds as part of an integrated approach is in scope of this book chapter.

## 1.1 Requirements Engineering and its limitations

Many companies struggle in the complex endeavour of establishing a high-quality Requirements Engineering and, in consequence, many projects suffer from insufficient Requirements Engineering. One of the key characteristics of Requirements Engineering is its volatile nature and its sensitivity to its practical context. Many things are not clear at the beginning of a project and a methodology, method, or tool that might fit very well the needs of one project could be completely alien to the needs and the culture of the next. This is what renders Requirements Engineering as something hardly standardisable with universal one-size-fits-all solutions and, thus, a discipline difficult to master. It is therefore not surprising that 33% of software development errors are estimated to have their origin in insufficient RE (Emam and Koru 2008; Méndez Fernández and Wagner 2014). Moreover, 36% of those errors are known to lead to project failures. Requirements Engineering is therefore not only difficult to handle, but it is also crucial for project success. Further studies corroborate the criticality of Requirements Engineering as they show how requirements errors may represent 40% of the total project costs; it is commonly accepted that when these errors are found late in the development process, their correction can make up to 200 times more than when correcting them during in early development stages (Venkatesh Sharma and Kumar 2013).

---

[1] See the IREB glossary, available at www.ireb.org

[2] We can often observe that RE is subsumed under the umbrella of software process models or product management approaches, often without using the term "Requirements Engineering". In this chapter, we do not distinguish between those various approaches but refer to the handling of requirements – from their inception to their specification and validation – which is in scope of any product development regardless of the chosen approach and terminology and regardless of whether it is done explicitly or implicitly.

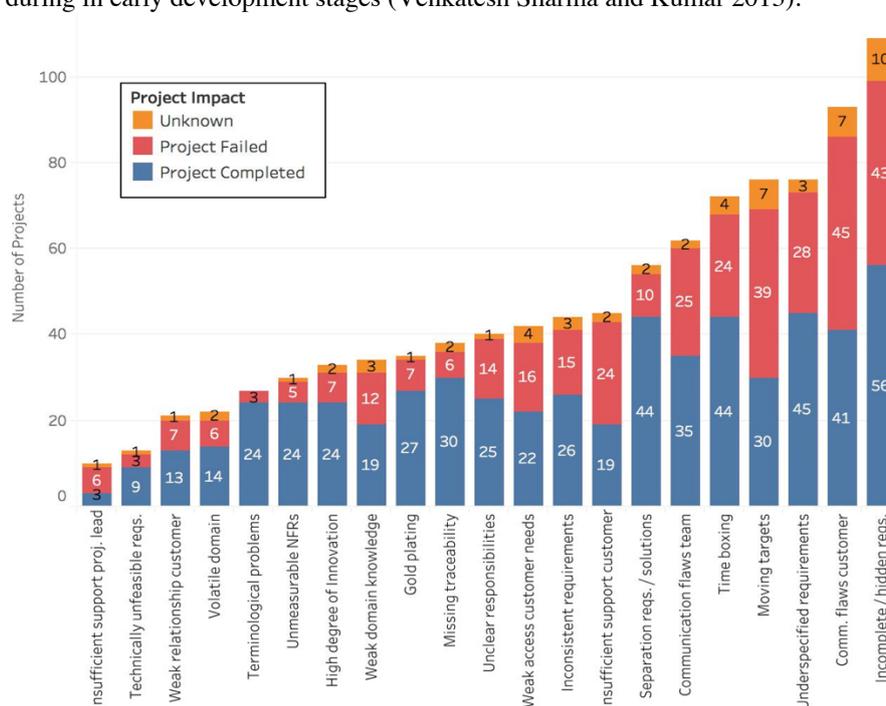

**Figure 1** Top 21 Requirements Engineering problems as revealed by the "Naming the Pain in Requirements Engineering" initiative. See also www.napire.org for more information including publications and open data sets.

We initiated, at the time of writing this chapter already a decade ago, a globally distributed, bi-yearly replicated family of surveys to gain insights into contemporary practices and challenges in Requirements Engineering: The Naming the Pain in RE initiative (short: NaPIRE, see also Wagner et al. 2019, Méndez Fernández et al. 2016)[3]. Among the insights we gained are a clearer understanding about the most frequently occurring and most critical problems companies experience, as well as their root causes and their effects (going beyond a binary view on project failure and success). Here, we discovered that a large share of problems is related to human factors and the lack of expertise to deeply penetrate the problem space – and this is regardless of the software process model employed such as "agile" (Méndez Fernández et al. 2015).

---

[3] See also the project website NaPiRE.org for further information and related empirical data sets.

This is not surprising given that Requirements Engineering is historically grown out of engineering disciplines and corresponding worldviews and it involves many different approaches, methods, tools, and techniques – none of which is suited for all purposes. In any case, while most of primarily academic debates are centred around questions related to the specification and refinement of requirements to measurable and, in particular, verifiable requirements covering various forms of representation (for models and natural language descriptions) as well as questions related to facilitating seamless modelling and the transition to the solution space in engineering, little attention is paid to eliciting the actually relevant requirements to obtain a sufficiently complete and correct requirements specification.

In fact, one of the biggest lies we tend to tell ourselves in Requirements Engineering is that the information relevant to understand the problem space (stakeholder information, context information, requirements) is omnipresent and simply needs to be elicited. A typical consequence of this problem is what we call "solution orientation" (Mendez Fernandez et al. 2012), the tendency of moving too fast to developing a solution and of focussing on related technical aspects often without a proper understanding on the problem to be solved by that solution. Here, actual user needs are often neglected, requirements are invented based on incorrect assumptions or blindly reused from other supposedly similar projects and solely based on the requirements engineer's intuition, or they lack creativity (see Inayat et al. 2015 as well as the results of the NaPiRE initiative). This underlines the need for more problem-oriented ways of thinking.

In fact, today's complexity growth in product development where system and domain boundaries become more and more fuzzy and where human factors become more and more important makes explicit the need for a shift even in RE itself (and corresponding roles and responsibilities) from often technology-centric ways of thinking, tasks, and domain-expertise to problem-centric ways of thinking, mediation, empathy, and creativity. This gives raise to the need for new approaches in interdisciplinary team configurations. This is what is promised by Design Thinking. But how does Design Thinking delineate exactly from Requirements Engineering?

## *1.2 Design Thinking and Requirements Engineering: Two distinct, yet complementary Approaches*

With its growing relevance in agile software development, Design Thinking has gained recognition as a creative problem-solving method, particularly when the real-world problem is complex or "wicked" (Buchanan 1992). Industry studies have highlighted this significant development. For example, based on a survey of the Hasso-Plattner Institute (Schmiedgen et al. 2015), over 69% of Design Thinking practitioners and managers identified Design Thinking as one of the major contributors to conduct an efficient innovation process. In a survey of IBM by Forrester (2018), Design Thinking was reported to reduce development and testing time by 33%, equating cost savings of around $1.1 Mio per major software development project. Some researchers even consider Design Thinking a "modern form of requirements engineering" (Beyhl and Giese 2016, p. 288) addressing some of the aforementioned challenges in current Requirements Engineering practices. However, we argue that this is not the case. Design Thinking and Requirements Engineering emerge from different backgrounds and offer different tools and approaches aiming at different goals, even though these goals are complementary by nature, as explained next.

In principle and as elaborated in more detail in the next sections, when developing a software-intensive product, we need to accommodate essentially two perspectives. On the one hand, we need a profound understanding of the socio-technical and the operational context of the system under consideration. It is important to elaborate what problems, needs, and goals stakeholders really have, and what the particularities of the domains including limiting (e.g., regulatory) constraints and demands are. This constitutes the difficult task of gaining a profound understanding of the too often fuzzy goals, rather than requirements or solution proposals, what their implications are, and what possibilities these goals open for future products. On the other hand, we need to elaborate a solid foundation for the engineering of a software product where we clearly specify – as far as possible in a non-technical, solution-independent fashion – what the elementary functional and non-functional properties of the software product are. Those properties build the basis for a variety of engineering and management activities ranging from architectural design over implementation and verification to project organisation and planning activities such as effort and cost estimations.

The first perspective is what is typically in scope of Design Thinking which describes a specific mindset and often non-technical approaches to penetrate the problem space from a user perspective and to deliver non-technical throw-away prototypes that allow to better understand that problem early on. The second perspective is what is typically in scope of Requirements Engineering which describes (engineering) methodologies, approaches, and tools to specify requirements in a detailed and testable way that facilitates subsequent development and management activities in a seamless manner. Here, capturing the problem domains and user perspectives is in many ways important (think for example in terms of UX), but not always central. A central task in Requirements Engineering is often to focus on operational environments and underlying infrastructures as well as their technical constraints, implications, and cost structures, but also on evidently demonstrating compliance to regulatory standards existing for many industries (think for example in terms of safety-critical systems or cyber-physical systems). In that sense, Requirements Engineering comes in many forms and interpretations which are all different to the principal ways of working in Design Thinking and yet they are all complementary to each other.

Design Thinking leverages interdisciplinary teamwork for a structured approach of ethnographic methods, and fast and simple (non-technical) prototyping cycles to produce innovative solutions in early product, service, and system development processes (Brown 2008; Kolko 2015). This rather diverging nature of problem-solving is notably different from the more converging ways of Requirements Engineering practices in most software-intensive projects (Harte et al. 2017). The multi-faceted opportunities of applying Design Thinking for Requirements Engineering are highlighted by several research community members. Vetterli et al. (2013) were ones of the first who suggested bringing Design Thinking and Requirements Engineering together for developing software applications. Academics with a content-focused view (*what* value does Design Thinking add) have recognized its value in terms of product quality, user acceptance, and process speed, mostly in specific domains like learning environments (Soledade et al. 2013), social innovation (Newman et al. 2015), or health care (Harte et al. 2017). Academics with a more process-focused view (*how* does Design Thinking add value), examine usage schemes of Design Thinking with software engineering techniques and agile development toolkits. For instance, authors have investigated the integration of Design Thinking and Scrum (e.g., Häger et al. 2015; Przybilla et al. 2018) and found evidence for higher innovation potential stemming from a combination of both approaches. Although mainly practice-oriented literature suggests potential benefits of combining Design Thinking and Requirements Engineering, or more generally speaking Software Engineering, knowledge on how this could be done in a seamless manner remains still unclear (Beyhl and Giese 2016).

While Requirements Engineering is a rather mature discipline with a long-standing history in research and practice, resulting in a plethora of holistic methodologies, practices, and tools, there is still limited knowledge about Design Thinking as Yoo (2017, p.v) emphasizes in his call to "advance the intellectual foundation of Design Thinking" for IS research. Little is known, in fact, about the specific impact on Requirements Engineering. A deeper understanding of Design Thinking would enable both communities, Requirements Engineering and Design Thinking, to evaluate its application purpose and potential for discovering and specifying requirements more thoroughly.

This is what is in scope of this chapter in the hope to provide a solid foundation for the remainder of this book.

## 1.3 Contribution and Outline

In this chapter, we elaborate on an effective integration of Design Thinking into Requirements Engineering. Note that we do not pretend that there would be one exclusive way of doing Design Thinking or Requirements Engineering. Rather, our aim is to introduce the mindset and common practices of both worlds, abstract from those practices by means of concentrating on the underlying outcomes (artifacts), and finally to use those resulting more simplified models for an integration of Design Thinking and Requirements Engineering which we further complement with practical experiences and recommendations. This provides a common basis for the various invited expert discussions captured in this book.

In the remainder of this chapter, we focus on the following contributions:

- First, we introduce the very fundamentals of Design Thinking and Requirements Engineering including the principles and practices as often found in literature.

- We then elaborate a first artifact model for Design Thinking that captures the essential concepts, approaches, and terms, and we will do the same for Requirements Engineering. We particularly concentrate on an artifact-centric view as a process-agnostic means that allows us to concentrate on the essential work products and their dependencies while abstracting from the particularities of surrounding, often very complex and unique specific-purpose processes.

- We use the artifact models for Design Thinking and Requirements Engineering to propose an integration of both.

- To use that integration not only as a conceptual foundation but also allow for its effective use in practice, conclude by introducing different operationalisation strategies on how to make efficient use of the introduced combination of Design Thinking and Requirements Engineering to create human-centred software-intensive systems.

Rather than merely focusing on presenting an academically oriented concept model, we aim at elaborating on essential terms, principles, and concepts while considering (and extending) the perspective on the practical relevance as many results emerge from academia-industry collaborations.

One central hope we associate with this introductory chapter is therefore to set the foundation for the subsequent invited chapters and to contribute to the ongoing debates and efforts in effectively integrating both worlds.

## *1.4 Previously Published Material*

Note that the insights provided in this book chapter emerge from previously published material, among it the dissertation of the first author (Hehn 2020) as well as the long-term collaboration with the second author. In some parts, we will explicitly borrow from parts of the dissertation in a verbatim manner.

# 2. Conceptual Background

In the following, we introduce the background to the extent necessary for the contributions of this book chapter. We will first elaborate on the very fundamentals of Design Thinking before concluding with a brief introduction of Requirements Engineering.

## *2.1 Design Thinking as a human-centred problem-solving approach*

Design Thinking is referred to as "a human-centered approach to innovation that draws from the designer's toolkit to integrate the needs of people, the possibilities of technology, and the requirements for business success." (Brown 2012) The roots of Design Thinking date back to the late 1960s, when design academics examined the mental processes that underlie design activities and transformed them into normative guidelines for creative problem solving (Simon 1969). These studies have expanded the scope of design beyond the boundaries of product styling to a way of thinking that can now be universalized for a multitude of disciplines (e.g., management, business, software development, engineering).

The paradigm of human-centred design is both starting point and foundation of all activities at all stages in Design Thinking (Brown 2008). Design Thinking solutions evolve from the triad of human values (desirability), technological feasibility, and business viability, combining expertise from the field of design, ethnologic and anthropologic research, engineering, and business economics. The dimension of desirability (what people want and need) anchors in a deep empathy for users and is applied by involving relevant stakeholders systematically throughout the entire process. Diverse design techniques help facilitating the creative transformation of user knowledge and insights into new concepts. Subsequently, feasibility and viability are integrated and explored. The lens of feasibility (how technology can help), therefore, demands an exploration of organizational capabilities and technological options in order to translate the human-centred requirement into actual products and services. Assessing the third dimension of business viability (what is financially sustainable) entails evaluating market opportunities and their compliance with the business objectives of the organization. Given its integrative nature, Design Thinking can be applied to "all aspects of business and society" (Brown 2009, p. 3) and is equally relevant for designing tangible and intangible solutions, both in public and private sectors.

## 2.2 Design Thinking on an operational level

On an operational level, Design Thinking is interpreted in three ways: as (1) a process with a sequence of steps according to a prescriptive process framework, (2) a toolbox with a collection of methods for situational support, and (3) a mindset with a set of human-centered principles to be internalized (see Figure 1). While all three modes are interlinked, they result in different conceptualisations on a practical level. As Fraser (2011) suggests, "it takes a combination of the right mindset (being) and a rigorous methodology (doing) that unlocks a person's thinking, and that one must consider all three of these factors." (p. 71)

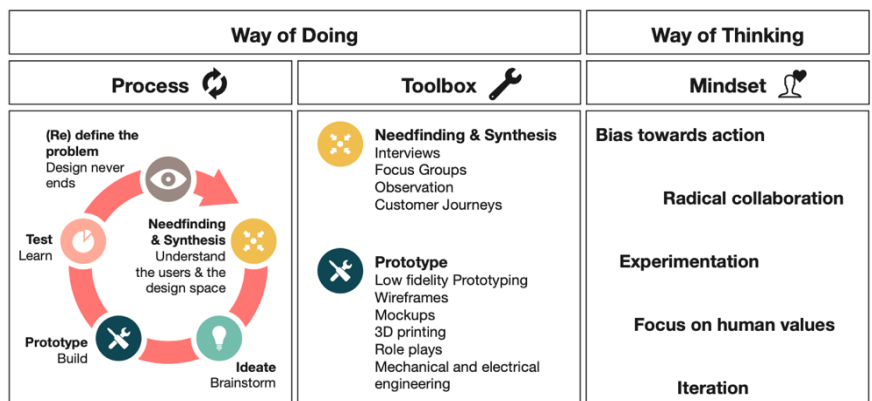

**Figure 2** Design Thinking as Process, Toolbox, and Mindset (see also Hehn 2020)

### 2.2.1 Design Thinking Process

Design Thinking process models, accompanied by a set of design tools, provide a supportive framework for practical use. Because of their specific character and clear instructions, those models are often utilised in Design Thinking education to provide a tangible, formalized approach to the Design Thinking concept. The normative Design Thinking process model is typically divided into two main phases: (1) problem exploration with problem definition, needfinding, and synthesis and (2) solution exploration with ideation, prototyping, and testing (ME 310 2010).

The Design Thinking process as introduced in the left column in Figure 2 can be summarised in five iterative steps as illustrated in the following:

1. *(Re-)Define:* The starting point of Design Thinking process implies an intensive level of engagement with the topic under consideration. The complex problem is transformed into a single sentence (often starting with "How might we...?" or "What if…?") entailing a clear design challenge and, therefore, making the topic somewhat tangible. Activities in this stage are to identify sources of inspiration, assess relevant stakeholders and their impact on the problem, explore emerging trends and market adjacencies, and to prepare research directions.

2. *Needfinding & Synthesis:* In the second step the topic is concretized by collecting user data through field research. The design thinking team applies empathic research techniques to uncover hidden needs and unexpressed desires by finding out how people work, what they like and dislike, and how they interact with a product or service. Practical activities in the observation phase include interviews (e.g., with users, extreme-users, non-users, and experts) and a variety of observation methods (e.g., self-documentation, on-site observation, shadowing). The acquired needfinding data is then transformed into meaningful insights about (unmet) user needs. Problem framing and reframing helps to identify patterns and ultimately develop a focus on where to create the highest value and impact for them. Applied tools are storytelling, scenarios, empathy maps, journey maps, and personas.

3. *Ideation*: Based on the developed insights in step 2, structured creativity methods support idea generation for new solutions. Ideation focuses on creating ideas and concepts (for instance by brainstorming techniques) as well as sketching them out quickly. Brainstorming rules such as "be visual", "encourage wild ideas", "defer judgment", "go for quantity", "stay focused on topic", and "build on the ideas of others" are applied to stimulate creativity and thinking outside the box.

4. *Prototyping*: Solution ideas that seem promising are turned into tangible prototypes (e.g. (paper-) models, mock-ups, role-plays, storyboards, journey mapping, short videos) in order to facilitate communication and feedback from end users. Therefore, it is not necessary to build perfectly well-engineered products, but rather simple versions and multiple alternatives in parallel, which focus on the most important aspects or highlight features for which feedback is crucial. Over the course of a project, prototypes usually evolve from so called 'Critical function/experience prototypes' (that define the core functionalities of the solution), and 'dark horse prototypes' (that challenge key assumptions and boundaries with visionary ideas) to 'system prototypes (that combine the most promising elements into one system vision (Uebernickel et al. 2015).

5. *Testing*: The ultimate step is the collection of user feedback and definition of improvement opportunities. Since it is important to understand the physical interaction of the product in use, feedback from end-users and project stakeholders is processed for further concept enrichment and revision. Considering the new information, the Design Thinking team may then go back to earlier steps, often revising the point of view stage or even starting the entire process over again by doing additional research about a specific idea and its realization.

Another way of visualizing the innovation workflow is dividing the Design Thinking process into two exploration stages: (1) the exploration of the problem space and (2) the exploration of the solution space, both consisting of an interaction between information gathering (divergent activities) and information processing (convergent activities). This visualization is also called "Double Diamond" (see Figure 3).

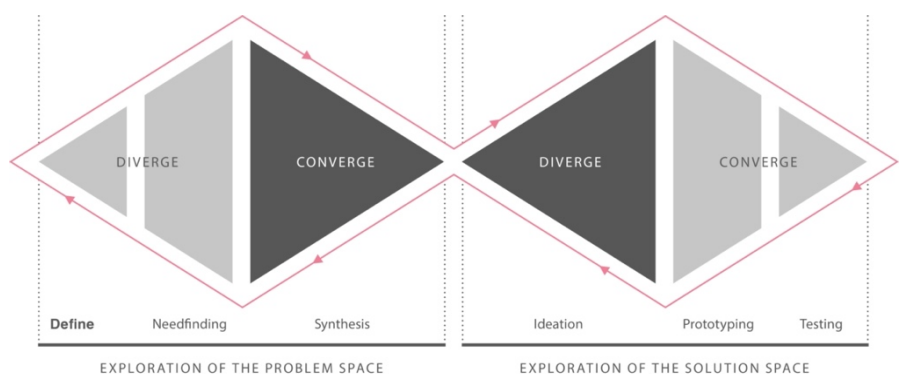

**Figure 3** Double Diamond (see also Hehn et al. 2020, p. 26)

The problem space demands diligent examination of the problem context by integrating all relevant stakeholders and the synthesis of all collected information to a clearly defined point of view, including needs and insights. The solution space encourages the generation of ideas and the creation of prototypes, which can be evaluated and tested with users. The process is repeated several times until a final solution can be presented. Reflection points are carried out during the process wherever necessary as they are crucial steps for adapting to novel information and developing deeper insights. Each cycle stimulates creativity and encourages rapid learning through trial-and-error.

### 2.2.2 Design Thinking Toolbox

Design Thinking as a toolbox breaks Design Thinking down into a set of techniques from which to pick and choose those that work best for the particular context and situation (see middle column in Figure 2). A wide range of practitioner catalogues of Design Thinking methods and tools have emerged in recent years (Doorley et al.

2018; IDEO.org 2015; Uebernickel et al. 2015). In this case, Design Thinking is not so much considered a prescriptive process or a distinct phase of a process, but rather a bundle of handy and selective (design) methods and techniques for situational support. Examples of the most used methods that are attributed to Design Thinking are summarized in the following table (Hehn et. al 2018).

**Table 1** Examples of Design Thinking Methods (adapted from Hehn et al. 2018)

| Method | Description | Phase |
|---|---|---|
| Stakeholder Mapping | Analysis of all stakeholders that are affected by the design challenge | Define |
| Desk Research | Desk research is known for collecting data based on literature and internet research | Define |
| Framing & Reframing | Framing and reframing is used to define the scope (and out of scope) of a project | Define |
| Interviewing | Conversation between two or more people where questions are asked by the interviewer | Needfinding |
| Observation | Observation and descriptions of happenings in the real world | Needfinding |
| Active Listening | Technique to elicit needs by understanding and responding to what someone has said | Needfinding |
| Clustering | Technique to bundle ideas and statements into thematic buckets | Synthesis |
| Storytelling | Method for exchanging knowledge collected during needfinding | Synthesis |
| Insight Formulation | Processes to distill and capture the most important learnings from needfinding | Synthesis |
| "How might we" Questions | "How might we …" is a way of asking questions to initiate Brainstorming but also entire projects | Ideation |
| Brainstorming | Brainstorming is a group creativity technique, mostly based on Osborn's method | Ideation |
| Brainwriting | Similar technique to Brainstorming but all ideas are collected in written format before the information exchange within the group starts | Ideation |
| Paper Prototype | Tangible representation of a product or service to facilitate testing | Prototyping |
| Role Playing | Role playing is used to act out service scenarios quickly and simply | Prototyping |
| Sketches / Scribbles | Sketching and Scribbling is all about drawing ideas and making them more tangible. | Prototyping |

| Feedback Capture Grid | Framework to capture user, customer, or stakeholder feedback while testing (dimensions often are: Likes, Criticism, Ideas, Questions) | Testing |

Contrary to the process-perspective, the toolbox offers an even more flexible way of using Design Thinking and tailoring it to specific project conditions. Thus, it can be integrated into the daily work routine and into existing company structures relatively quickly. However, since many of the Design Thinking techniques are not necessarily exclusive to this approach, it may raise the question from which point onwards to actually speak of Design Thinking.

### 2.2.3 Design Thinking Mindset

A growing number of authors stress that the core of Design Thinking goes beyond process models and tools (e.g., Kröper et al. 2010; Martin 2009). They perceive Design Thinking primarily as a mindset or general "design attitude" (Boland & Collopy 2004, p. 3) towards creative problem-solving (see right column in Figure 2). This entails the development of empathy, an open-minded and optimistic approach to generating insights and ideas, and the rationality to investigate and fit those ideas in compliance with the context. The main principles are highlighted in the following:

- *Design Thinking emphasizes human values as a starting point and foundation for all related activities* (Brown 2008). Understanding what people need and want anchors in a deep empathy for users and is achieved by systematically integrating a variety of stakeholder groups throughout the development process, both through direct dialog and non-obtrusive observation methods.

- *Design Thinking solutions are mainly generated through radical collaboration*, both with users and by composing a multidisciplinary project team that incorporates different functions and departments (Doorley et al. 2018). By encouraging inter-organizational cooperation on the ground of common principles for a collaborative culture, Design Thinking can be regarded as a holistic framework for co-creation.

- *Design Thinking leverages abductive reasoning to constantly generate new information and consider alternative options early on*. The abductive nature of this way of working induces a "reflective conversation with the situation" (Schön 1984, p. 76) by looking beyond "what is" and exploring the logic of "what might be" to generate customer and business value (Martin 2009).

- *Design Thinking stresses a bias toward action*. This means that the preferred ways for gathering insights and feedback from stakeholders are hands-on activities such as experimenting with ideas, building prototypes, and testing them (Doorley et al. 2018).

- *Design Thinking is a fundamentally exploratory process that encourages rapid and iterative learning cycles.* According to the "fail early and often"-principle every iteration leads to further adjustments and new directions in the development process. In the long run, this iterative approach to development is supposed to mitigate risks of not meeting customer needs in the long run (Brown 2009).

## 2.2 Artifact-based Requirements Engineering and the AMDiRE Approach

Similar as it is the case for Design Thinking, Requirements Engineering, too, comes in various forms and interpretations while none is best for all purposes. In this chapter, we will not even try to introduce the discipline in its various interpretations to the extent they deserve, same as it is not our intention to promote any of the various (and often competing) approaches to Requirements Engineering. Rather, we aim at laying the foundation for an Requirements Engineering that integrates the very Design Thinking tools and principles introduced above.

In principle, how Requirements Engineering is done in practice – including the artifacts created and the techniques used – depends on many factors such as surrounding software process models, application domains, industry sectors, and even engineering cultures including personal, subjective preferences in engineering terms. Those characteristics render Requirements Engineering approaches as something unique and barely standardizable with a one-size-fits-all solution. In response to this complexity in the choice of methods and approaches various artifact-based approaches to Requirements Engineering have been elaborated over the last two decades. All those approaches capture the particularities of the envisioned domains and serve as reference models to guide the elaboration of precise requirements for those domains while offering the necessary flexibility in how to do it from the perspective of processes and activities. To this end, there are several blueprints of the results and their dependencies rather than a dictate for complex activities, tasks, or methods. This is what essentially reflects the artifact-centric philosophy. In such a philosophy, we concentrate on defining the artifacts, their contents, and their dependencies in a central model that constitutes the backbone of a (Requirements Engineering) project, and which leaves open when to create which artifact and which description technique to use (Méndez Fernández et al. 2019). Such a model then serves as a guidance for engineers in elaborating their results (e.g., the specification of user requirements via use cases and capturing their relationship to acceptance test cases to support traceability) while leaving open how they intend to do it (e.g., in an agile manner or a rather plan-driven manner).

In this book chapter, we rely on one specific artifact-based approach to Requirements Engineering which we use as integration point for Design Thinking. The approach we rely on is the Artifact Model for Domain-independent Requirements Engineering (short: AMDiRE). AMDiRE emerges as a concluding synthesis of the various approaches developed in recent years for different domains and industry sectors and which all have been disseminated into everyday practice, e.g., at Capgemini, Siemens, Bosch, BMW, or Cassidian. The AMDiRE approach thus emerged as the result of consolidating previously developed approaches and the lessons we learnt during their development, evaluation, and dissemination.

Here, we focus on the very foundation of AMDiRE to the extent necessary in the context of our chapter. Details on the approach can be taken from previously published material (Méndez Fernández and Penzenstadler 2014).

### 2.2.1 Overview of AMDiRE Components

Figure 4 illustrates all components included in the AMDiRE approach necessary to use it as reference at project level. The central component of AMDiRE is defined by its artifact model. For the sake of simplicity, we see an artifact as a key deliverable of major interest that abstracts from contents of a specification document. It can be used as input, output, or even as an intermediate result in Requirements Engineering created along a particular task or method and by choosing a description technique (e.g., natural language, structured tables, figures, or models) as long as it complies with the artifact model as explained next. A more insightful introduction into what artifacts are in software engineering can be taken from our reflection provided in previous work (Mendez Fernandez et al. 2019).

For each artifact, we capture two essential views: a structure view and a content view. The structure view captures for each artifact type (e.g., "requirements specification") the content items to be considered (e.g., "use case model"). For each content item, we define the content view via the modelling concepts, e.g., the elements and content relations of a use case model and different description techniques that can be used to instantiate these concepts, such as an UML activity diagram. The structure model thus gives a simplified view on the content and is used to couple the contents to the elements necessary to define a process, e.g., roles, methods, and milestones relevant for a use case model. This is in scope of the integration of AMDiRE into company- and project-specific software process models (often referred to as static tailoring). The content model then guides by defining what is necessary to specify the content, e.g., scenarios, actions and actors, which we use to create a use case model.

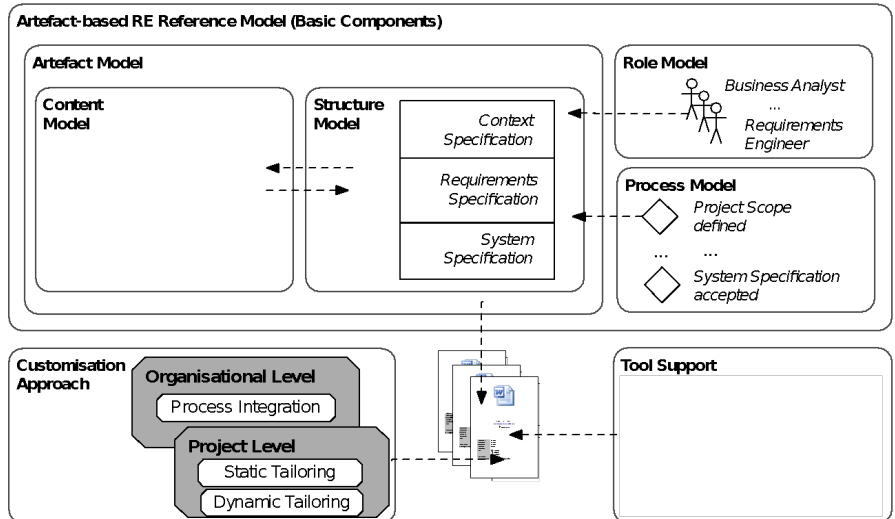

**Figure 4** Overview of AMDiRE Components (see also Méndez Fernández and Penzenstadler 2014)

Note that we consider – same as for activity-centric approaches to Requirements Engineering – elements of a process description, but instead of defining the process based on phases, activities, and methods, we define the process based on the artifacts to be created and their relationships, as well as related milestones for when these artifacts should be of sufficient quality to specify the next. Even though the content model supports the precision of the results in the flexible process definition, the process itself remains undefined. Regarding the methods and description techniques for creating the contents (e.g., UML or natural text), we leave open which one to choose, as long as the contents and relationships proposed by the artifact model are specified.

## 2.2.2 AMDiRE Artifact Model

The AMDiRE artifact model comprehends concepts used to specify the contents of the artifacts over three levels of abstraction: the Context Layer, the Requirements Layer, and the System Layer. Each of those levels of abstraction features a specified number of content items that are detailed in concepts used for a stepwise refinement of the various (modelling) views we have on a system. The context layer considers the context of a system, i.e. the domain in which to integrate the system such as the business domain with the business processes to be carried out. The requirements layer considers the system from a black-box perspective. That means, we specify the requirements on the system and the user-visible functionality from a perspective in which the system is intended to be used, without giving details about its technical, internal realization. That view is captured by the system layer which provides the glass-box perspective on the internal (logical and technical) realisation of the system.

The artifact model is in the center of our attention and consists of two basic models: the content model and the structure model. The content model abstracts from the modelling concepts used for a particular family of systems in a particular application domain over the defined levels of abstraction. The structure model gives a logical structuring to those concepts and is used for the integration with the role model and the process model (see also the previous section).

Finally, details on the single content items as well as further components which accompany AMDiRE will be introduced in context of the integration of our Design Thinking model into AMDiRE (while also referring the interested reader to the main article Méndez Fernández and Penzenstadler 2014).

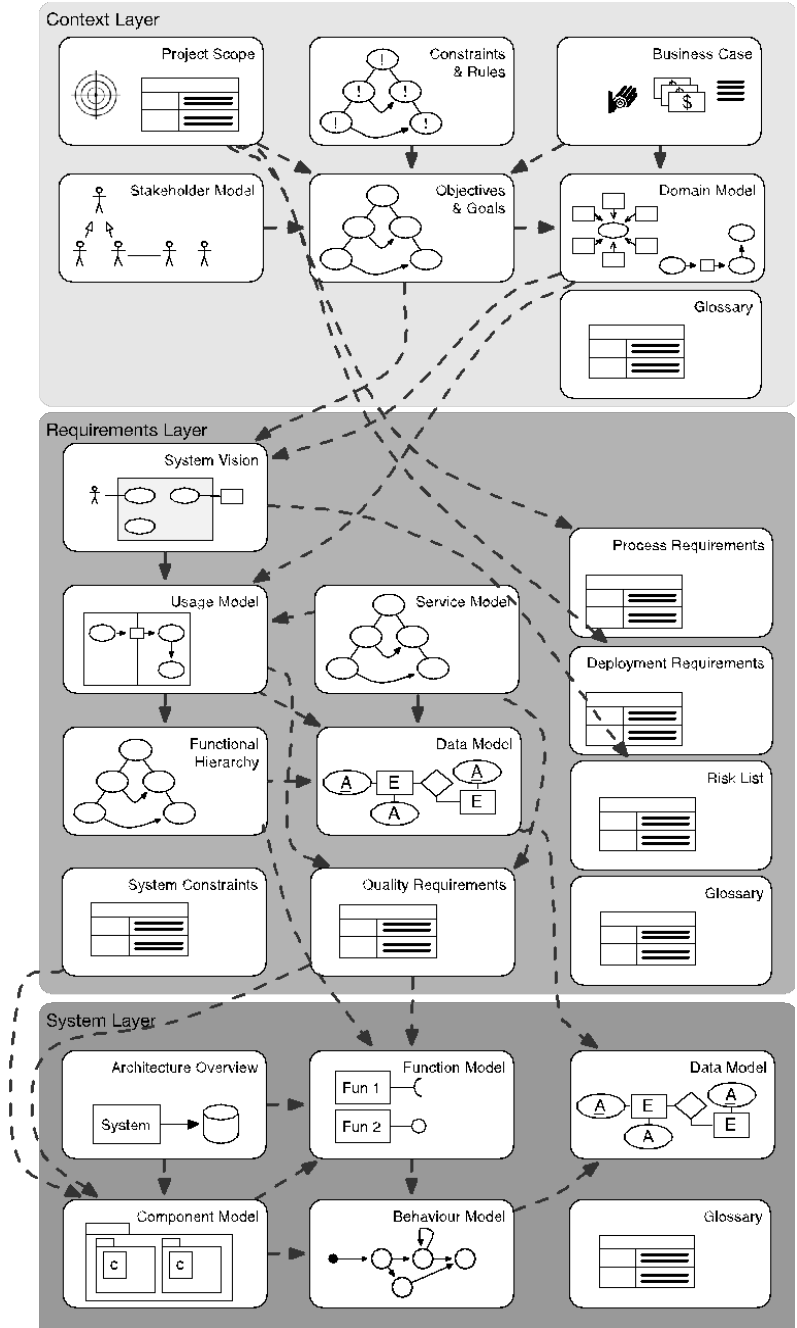

**Figure 5** AMDiRE Artifact Model (simplified view on structure model, see also Méndez Fernández and Penzenstadler 2014)

## 3 An Artifact Model for Design Thinking

In contrast to Requirements Engineering, no artifact model exists for Design Thinking – until now. We have taken the multitude of practitioners' compendia that present and summarize Design Thinking-specific methods as a basis to logically infer the results they produce (i.e., artifacts) (Gutzwiller 1994). Hence, we can rely on the available literature corpus as well as the knowledge we accumulated in our own more practically-oriented work as the foundation for determining, synthesizing, and summarizing the artifacts in a Design Thinking-based artifact model that is described in this section. Figure 6 presents the development steps we followed.

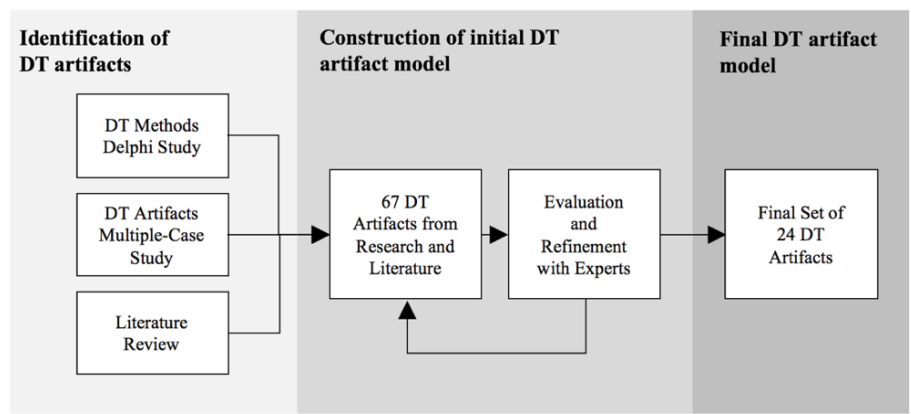

**Figure 6** Development Steps of a Design Thinking-based Artifact Model (see also Hehn 2020)

*Identification of Design Thinking artifacts:* Three sources of evidence provide data triangulation (and construct validity) to identify relevant Design Thinking artifacts (Yin 2014). The results of a Delphi study about the most used methods in Design Thinking (Hehn et al. 2018), empirical findings from multiple-case studies (Hehn & Uebernickel 2018; Hehn et al. 2018), and existing practitioner catalogues (Doorley et al. 2018; IDEO.org 2015; Uebernickel et al. 2015) serve as our main basis. The final set of artifacts included 65 Design Thinking-related artifacts.

*Construction and evaluation of an initial artifact-based Design Thinking model*: The initial model with 65 Design Thinking artifacts was evaluated in unstructured interviews with four Design Thinking experts from academia and industry. The experts were required to have either applied or researched Design Thinking methods for a considerable amount of time. Specifically, people were chosen when they had a proven track record of using Design Thinking in the context of innovative software-intensive projects for the past three years. Based on the feedback, three main findings evolved: First, the completeness of relevant artifacts and their attributions to the Design Thinking phases have been corroborated by all experts. Second, the original structure was adapted for better readability and comprehensibility from top to bottom according to the chronological order in which they typically appear in a project. Third, the model was refined to fit the frame of reference in terms of granularity of the artifacts. The second version of the model encompasses 21 artifacts and is presented in this book chapter.

*Construction of the final artifact-based Design Thinking model:* The revised and final version of the artifact-based Design Thinking model is visualized in Figure 7. It encompasses 24 Design Thinking artifacts structured into problem-oriented artifacts (sub-classified into define, needfinding, and synthesis) and solution-oriented artifacts (sub-classified into ideation and prototype & test).

A more detailed description of each content item can be found in the Appendix.

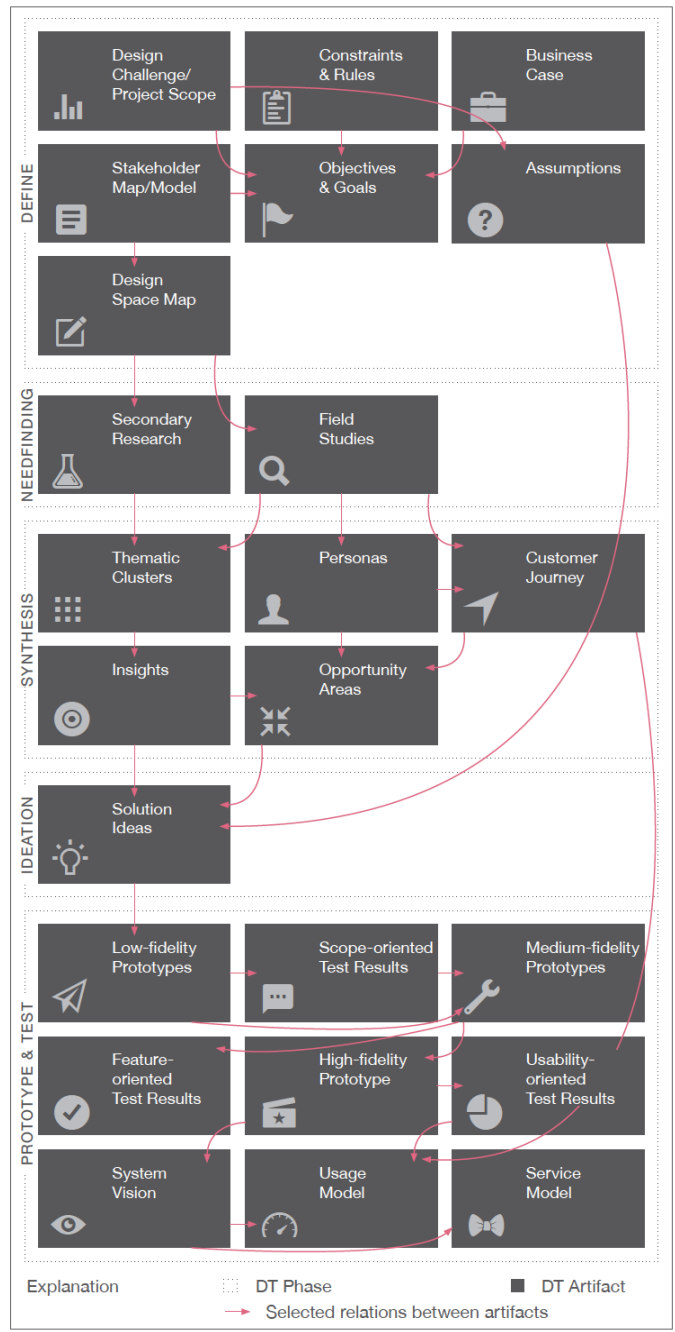

**Figure 7** Design Thinking Artifact Model (see also Hehn 2020)

# 4 An Integrated Artifact Model Combining Design Thinking and Requirements Engineering

In the following, we present an integrated model that combines Design Thinking and Requirements Engineering artifacts. We motivate the development, the structure, and implications for researchers and practitioners.

## 4.1 Development of an Integrated Artifact Model

An artifact-oriented reference model, such as those shown in the sections before, and that aims at integrating Design Thinking into a holistic engineering context is, as we argue, the only appropriate way to accommodate the variety of processes and methods of both approaches. Artifacts determine what must be accomplished (the work products and their interdependencies) instead of how it has to be accomplished (the steps that have to be taken). Further, defining a comprehensive view of the "desired" system and its key functionalities and features is an important objective of both Requirements Engineering and Design Thinking. The artifacts produced along Design Thinking and Requirements Engineering activities are used to support product design and project management decisions throughout the development process and product life cycle. The quality and appropriateness of these artifacts is therefore imperative for the successful development and acceptance of a software-intensive system. A model that encompasses the relevant artifacts of Design Thinking and Requirements Engineering can outline the synergies and differences between both approaches. While keeping a consistent structure and terminology, this condensed view focuses on the created work products, their contents, and dependencies, and it allows to abstract from their particularities of various processes and methods, which would otherwise render a comparison difficult.

Our integrated artifact model, therefore, contains and structures all the artifacts referenced, modified, or created in Requirements Engineering and Design Thinking in software-intensive development projects. To be useful, the model should support the re-use of knowledge and should be tailorable to certain situations in an efficient manner. The aim is to integrate Design Thinking and Requirements Engineering artifacts to simplify the adoption and configuration (i.e., usage schemes) of Design Thinking for Requirements Engineering.

Our goal was to establish a reference model that should
1. support the integration of both approaches respecting their different "flavours"
2. provide flexibility in the way of working to cope with the various influences in individual project environments and for organisational needs, and

3. enable a reproducible creation of work products in the context of innovative software-intensive development projects.

Similar as done for the development of the artifact model for Design Thinking itself, we show the steps for the construction and evaluation of our final combined artifact-based reference model for Design Thinking and Requirements Engineering in Figure 8.

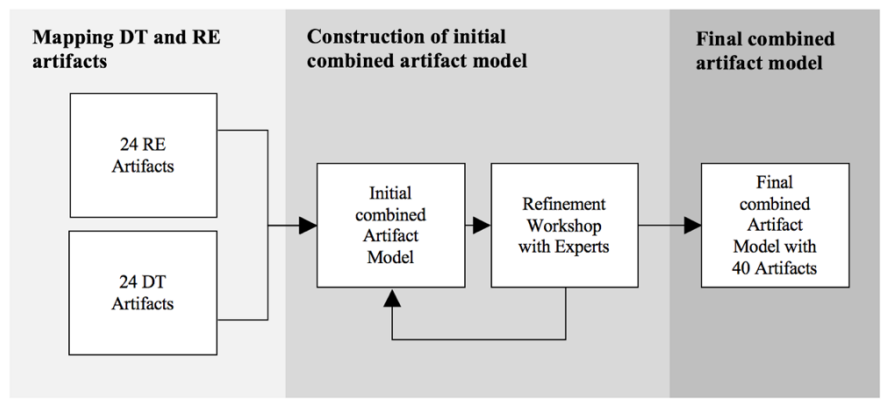

**Figure 8** Construction and Evaluation of an Integrated Artifact Model (see also Hehn 2020)

The process of mapping artifacts from Design Thinking and Requirements Engineering was performed by two experts in Design Thinking and Requirements Engineering. The comparison was performed with 24 Design Thinking artifacts and 24 Requirements Engineering artifacts. Based on these activities an initial integrated artifact model for Design Thinking and Requirements Engineering was created. This model has been continuously tested with Design Thinking and Requirements Engineering academics and practitioners to adapt the relevant artifacts and their interdependencies for a comprehensive overview. Details on the approach can be taken from Hehn 2020.

## 4.2 Integrated Artifact Model

The integrated artifact model is presented in Figure 9. It establishes a blueprint of relevant artifacts, i.e., the work results, contents, and dependencies of Design Thinking and Requirements Engineering. All artifacts are denoted in rectangles including the name of the artifact and a number. Associations depict relations between the artifacts, however not exhaustively, for reasons of reducing visual complexity. The Design Thinking phases (dotted line) provide a sub-structure for organizing the Design Thinking artifacts.

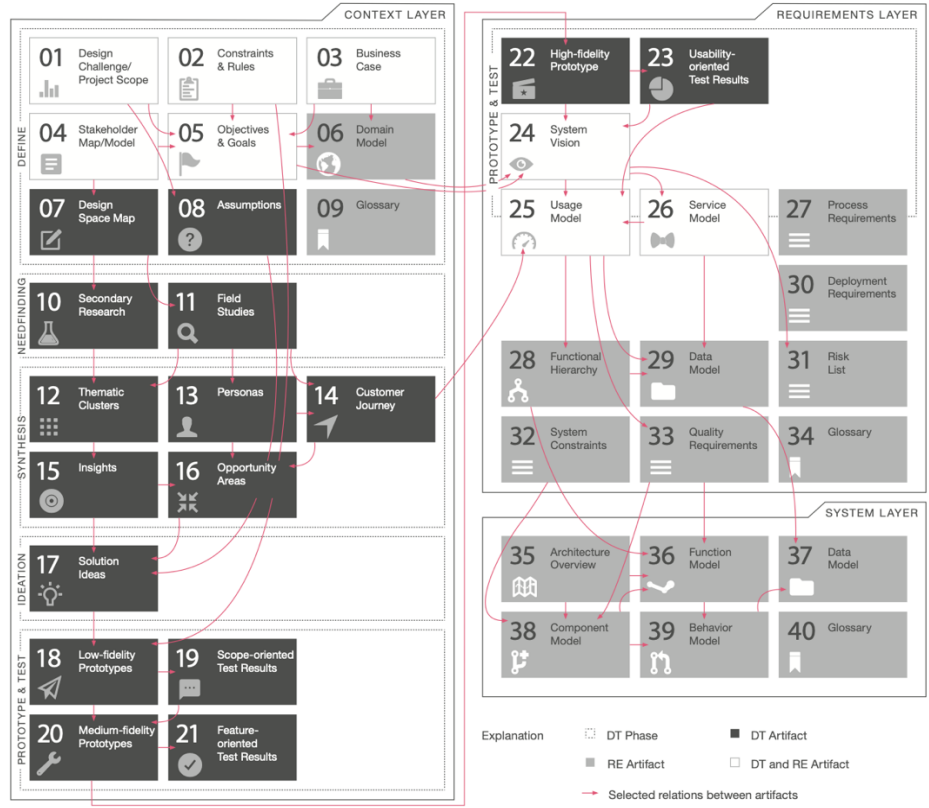

**Figure 9** Integrated Artifact Model (see also Hehn et al. 2020, p. 27)

Table 2 summarizes the elements used to compose the artifact model.

**Table 2** Overview of Elements in the Integrated Artifact Model

| Representation | Description |
|---|---|
| | The folder box denotes the layers context, requirements, and system as the overarching structure of the artifact model |
| | The dotted line indicates the Design Thinking phases (Define, Needfinding, Synthesis, Ideate, Prototype, Test) for means of comprehensibility |
| | The dark rectangle denotes a Design Thinking artifact including the artifact name, a number in the artifact model and an icon |
| | The grey rectangle denotes an Requirements Engineering artifact including the artifact name, a number in the artifact model and an icon |
| | The white rectangle denotes a combined artifact (Design Thinking and Requirements Engineering artifact) including the artifact name, a number in the artifact model and an icon |
| | The arrow denotes a unidimensional relation between artifacts. It expresses an input-output relationship |

The overall structure of the model is orientated along the three layers of the AMDiRE model (context, requirements, system) — each capturing a collection of relevant content items from Design Thinking and/or Requirements Engineering.

As discussed earlier, the *context layer* covers the information relevant to define the context and includes, for example, the overall project scope, stakeholder information, a domain model, and assumptions of the project team, and underlying goals and constraints. Hence, much of the information captured in Design Thinking concentrates on this layer.

The *requirements layer* encompasses what is necessary to operate in this context and captures, for example, the system vision, high-fidelity prototypes, a usage and behaviour model, and the function hierarchy as entry points for the system layer. Similar to the context layer, much of the information here is documented using natural language, occasionally reflected, however, also in models (e.g., for data and functional perspectives on user-visible system behaviour).

Finally, the *system layer* includes information on how the system is to be realized and includes, for example, a logical component architecture and a specification of the desired behaviour, e.g., via function models. Again, information within this layer is documented using both, natural language and conceptual models (data, function, behaviour).

The integrated artifact model consists of three artifact types that encompass 40 content items with various relations. Out of all content items, 16 can be associated with Design Thinking, 16 with Requirements Engineering, and 8 with both (see Figure 10). The latter can be further distinguished into artifacts with similar semantics but different purpose (3 out of 8). These include the design challenge/project scope (#01), the business case (#03), and the objectives and goals (#05). The main reason for their different purpose is that in Requirements Engineering these artifacts have a convergent nature while in Design Thinking they can be considered as open because they provide the opportunity for a broad context exploration.

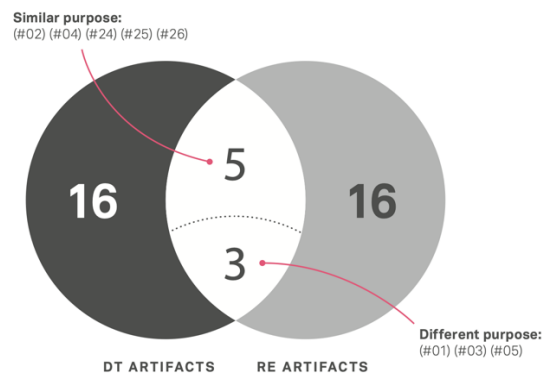

**Figure 10** Distribution according to Artifact Type (see also Hehn 2020)

The distribution of artifact types according to the specific layers in the artifact model is depicted in Table 3.

**Table 3** Distribution according to Layer (see also Hehn 2020)

| Layer | Design Thinking | Design Thinking and Requirements Engineering | Requirements Engineering | Total |
|---|---|---|---|---|
| Context | 14 | 5 | 2 | 21 |
| Requirements | 2 | 3 | 8 | 13 |
| System | 0 | 0 | 6 | 6 |

The model positions most artifacts within the context layer (21). Most Design Thinking-related artifacts can also be found here (14 Design Thinking only and 5 Design Thinking & Requirements Engineering artifacts). Next to the data model (#29, #37) the glossary (#09, #34, #40) is an Requirements Engineering-only artifact that can be found in all layers. This artifact type is revised based on the specific layer objectives. Starting in the context layer, the design challenge/project scope (#01) defines the relevant problem and primary scope of a project. Within this realm, the stakeholder map/model (#04) captures the most relevant stakeholders and their relationships. They provide one important rationale for the requirements and goals of the system (#05). The domain model (#06) contains context information and constraints (#02) about the operational environment connecting it to the requirements layer. Design Thinking artifacts complement and expand these mainly Requirements Engineering-related artifacts with a broad and human-centred perspective. For example, field study results (#11) and insights (#15) help to frame the project scope (#01) and inform specific use cases and scenarios (#25, #26) as defined in the requirements layer. Low- and medium-fidelity prototypes (#18, #20) are mainly leveraged to better understand stakeholder needs and system context.

The requirements layer contains five Design Thinking-related artifacts (two Design Thinking only and three Design Thinking & Requirements Engineering artifacts) and eight Requirements Engineering artifacts. The system vision (#24) denotes the general concept and idea of the intended system. High-fidelity prototypes (#22) are a way to visually enrich the system vision (#24) and to illustrate the key functionalities and general form of interaction (app, desktop solution etc.). Agreed upon by the relevant stakeholders, a system scope, i.e., major features and use cases as well as its constraints (#32), is specified. A service model (#26) defines the services the system shall offer complementary to the use cases defined through a use case model (#25). User-visible system functions are structured in a functional hierarchy (#28) which is the entry point into the system layer.

The system layer holds six Requirements Engineering artifacts and none of them are related to Design Thinking. While the context and the requirements layers include the information aspects that are typically found in Design Thinking- and Requirements Engineering-related artifacts, the system layer includes the items addressing what is known as the solution space and providing the interface for Requirements Engineering into design activities. In the system layer the functions of the functional hierarchy (#25) are related to components (#38), a functional model (#36), and their internal behaviour (#39), which also provides the basis to identify the data model (#37).

A more detailed description of each content item can be found in the Appendix.

### 4.3 Organizational Model

The integrated artifact model can be seen as a foundation for a more comprehensive organizational model that includes the following components: (1) the artifact model specifies what needs to be produced or exchanged; (2) the role model describes who should produce it and which particular responsibilities are needed; (3) the activity model describes what to do in order to create, modify, or use an artifact; (4) the process model denotes when the artifacts, roles, and activities should be produced or performed; and (5) standards and tools conceptualize with what all of the above mentioned activities are performed (Méndez Fernández and Penzenstadler 2014).

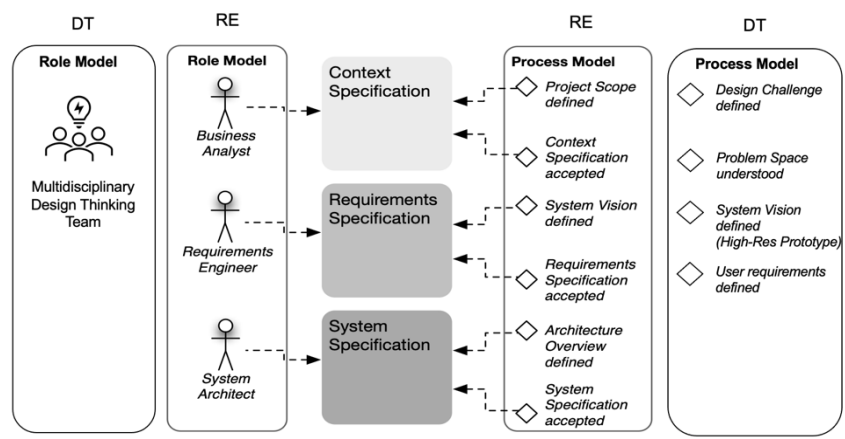

**Figure 11** Overview of Artifact Types, Roles, and Milestones

Figure 11 shows the artifact types in relation to roles and responsibilities (left side) and in relation to milestones (right side) which can be used to integrate the model into a process. We distinguish the Design Thinking- and the Requirements Engineering-view.

Note that in Requirements Engineering and in accordance with AMDiRE, we assign one role for each artifact type. Each role has the responsibility independent of other potentially supporting roles such as those provided by the surrounding software process model (e.g., product manager), and independent of whether same persons are assigned to different roles in a project. The Business Analyst has the responsibility for the context specification, the Requirements Engineer has the responsibility for the requirements specification serving also as a mediator between the business analyst and the system architect. That system architect, finally, has the responsibility for the system specification. In Design Thinking, a multidisciplinary team takes up the role to define the context and system vision. Often this team is drawn from various disciplines to integrate diverse perspectives constituting an important aspect in stimulating creativity and generating the potential for more comprehensive and original results. The willingness to cooperate with different people is an important aspect in Design Thinking practice since solutions are mainly generated through collaboration, both with users and by composing a multidisciplinary project team (around six team members). Typically, Design Thinking team structures are not subject to hierarchies and departmentalization but rather a way of radical collaboration that allows leadership to pass in-between members. Team members drawn from various disciplines integrate diverse perspectives constituting an important aspect in stimulating creativity and generating the potential for more comprehensive and original results. The versatile Design Thinker has acquired the position of a general problem solver possessing strengths in two dimensions which are commonly visualized as a "T-shape". Deep Knowledge corresponds to the academic expertise or a depth of skill that allows the Design Thinker to adapt their knowledge to the problem and make tangible contributions to the result. Broad knowledge and skills represent the ability to reach out to other specialists coming from a wide range of disciplines entailing a general openness to new ideas, people, and ways of doing.

For each artifact type, we furthermore define two milestones: An entry-level milestone indicates the point in time in which the first content item is expected to have a sufficient maturity in its content; for instance, the system vision in the requirements specification comprises an overview of the major use cases; its definition and agreement indicate that the use cases are succinctly defined to be further refined and modelled and, thus, allowing, for example, for first cost estimations based on function points. The second one indicates when the corresponding artifact is formally accepted.

Those milestones are sufficient for a process integration and instantiation as they give us the opportunity to formally embed the artifacts into project-specific decisions. Therefore, we enrich those existing milestones in analogy to the AMDiRE milestones to cover the Design Thinking artifacts following the same logic.

## 4.4 Findings and Practical Implications

Our integrated artifact model offers several important insights and implications for using Design Thinking in the context of Requirements Engineering. In the following, we highlight those we deem most important.

Various commonalities between Design Thinking and Requirements Engineering can be seen if the latter is understood as an iterative approach. The differences should be seen as complementary activities. The integrated artifact model distinguishes between more problem-oriented and more solution-oriented artifacts which addresses the principles of both Design Thinking and Requirements Engineering. Problem-oriented artifacts contain information about the underlying problem context including the goals and needs of stakeholders as well as specific system conditions or constraints. Solution-oriented artifacts contain information about the corresponding system vision and how to solve the problem stated in the project description.

The integrated model shows that Design Thinking mainly contributes to early Requirements Engineering activities with up to 14 additional context artifacts for a comprehensive understanding of the problem domain. Accordingly, Design Thinking expands the toolbox for Requirements Engineering by emphasizing the creation of artifacts that describe the relevance of the system vision. Design Thinking could even be exclusively used to perform these activities. A complementary approach of Design Thinking and Requirements Engineering, however, seems necessary for shaping the requirements layer. While both concepts produce overlapping artifacts (system vision, functional requirements, usage, and service models), their realization might take different forms. Design Thinking uses mainly a high-fidelity prototype to describe the system vision and key functionalities. Requirements Engineering specifies the same mainly by using rich picture and class diagrams. In addition, other requirement types, such as quality or deployment requirements are predominantly specified with common Requirements Engineering techniques. Requirements Engineering is exclusively used to specify system artifacts and to provide the interface to system design activities. Hence, Requirements Engineering also expands the toolbox of Design Thinking.

Following our AMDiRE role model as described in Méndez Fernández and Penzenstadler (2014) (see Figure 11), implications can be seen in expanding the knowledge of business analysts with Design Thinking skills and, vice versa, in equipping design thinkers with Requirements Engineering skills to gain appreciation for subsequent software design activities. Lauenroth (2018) calls this role 'digital designer' and defines them as "someone who is capable of creating a vision for digital products, processes, services, business models, or even entire systems, free from technical or organizational obstacles as well as apparent reservations (outside-in thinking). Digital designers are also capable of ultimately turning this vision into reality. They transfer (technological) possibilities into (new) product/process/service/business model/system design. To do all of this, digital designers must be skilled in design and the available technologies and be capable of interacting with all stakeholders." (p. 8) For training providers, the integrated artifact model can support the development of new training programs and learning formats about combining Design Thinking and Requirements Engineering. A new role with skills and talents in both approaches may be fostered. Current training courses in Design Thinking or Requirements Engineering can be enhanced by integrating the respective other approach to gain understanding about the benefits and shortcomings of the two incorporated concepts.

For project managers, several contributions can be seen. First, the model can be considered a support system to define and distinguish responsibilities in a project. Project roles can be directly coupled to the creation of artifacts, for which they must take the responsibility. Second, project managers can assign completion levels and establish progress control for the creation of artifacts. Quality assurance metrics can help to objectively measure the degree of completeness of an artifact in the artifact-based reference model. Third, the model ensures flexibility for integrating processes and customizing the reference model at project level. The combined model allows for variations of the created artifacts in response to individual project characteristics. For example, by defining the content-focus of the project, the creation of either Design Thinking or Requirements Engineering artifacts might be of greater help as each approach emphasizes a different content type. For example, to better understand the user and business context, the creation of Design Thinking artifacts might be preferred. Requirements Engineering artifacts should be at the center of attention to better describe the technical perspective and answer feasibility questions. Teams may also jump back and forth between both approaches if new questions come up in one or the other area. Fourth, the model can act as a basis for an effective requirements management, where the objective is to administrate the outcome of Requirements Engineering activities. This administration includes, for example, progress and traceability control, impact analyses, or risk mitigation (Jönsson and Lindvall 2005). A structured and consistent content specification is a prerequisite to perform such activities. Hence, the integrated artifact model can enhance the effectiveness of requirements management activities due to its defined set of interdependencies and chosen artifacts.

For team members of software-intensive projects (i.e., requirements engineers, business analysts, or design thinkers) the model offers a blueprint for creating syntactically consistent and complete results with respect to the respective application domain. While not all artifacts from the model must be considered in every project, the overview still serves as an orientation and connection to further design and development activities. The latter point is especially of interest for Design Thinking as this has been continuously criticized to be insufficiently linked to development processes (e.g., Häger et al. 2015).

# 5 Operationalization Strategies

In the following chapter we present three operationalisation strategies to integrate Design Thinking into Requirements Engineering when designing innovative software-intensive systems.

## 5.1 Overview

The integrated artifact model enables a flexible creation of the introduced Design Thinking and Requirements Engineering artifacts. This means that the decision which and when artifacts should be produced need to be customized according to specific project characteristics. To provide a guideline three operationalization strategies are proposed to integrate Design Thinking and Requirements Engineering in different ways. The strategies reflect existing research findings about integrating Design Thinking into software development practices (e.g., Dobrigkeit and de Paula 2019; Lindberg et al. 2012; Hehn & Uebernickel 2018).

We suggest the following three strategies: (1) Run Design Thinking prior to applying Requirements Engineering practices (upfront Design Thinking); (2) infuse the existing Requirements Engineering process ad-hoc with selected Design Thinking tools and artifacts (infused Design Thinking); or (3) combine the previous two strategies and integrate Design Thinking into Requirements Engineering practices on an ongoing basis (continuous Design Thinking). The ratio between Design Thinking and Requirements Engineering differs within the three proposed operationalization strategies. The better the original problem is understood, the more activities are biased towards straightforward design and implementation tasks (i.e., Requirements Engineering artifacts) (see Figure 12). The less it is understood, the more activities are directed towards context understanding and problem exploration (i.e., Design Thinking artifacts). Thus, the defined project objective and context are the guiding parameters for the selection of an appropriate operationalization strategy.

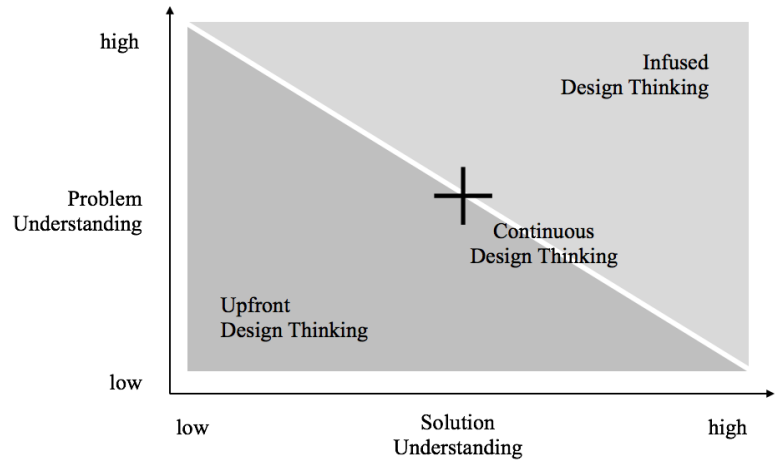

**Figure 12** Navigating Upfront, Infused, and Continuous Design Thinking Strategies (see also Hehn 2020)

## 5.2 Three Strategies to Operationalize and Integrate Design Thinking

In the following, we introduce our three strategies to operationalize our integrated Design Thinking approach. For each, we follow a structured approach of listing objectives, prerequisites, key activities, necessary roles, and outcomes followed by showing an exemplary practical case. This shall make our strategies more tangible.

### 5.2.1 Upfront Design Thinking

*Objective:* Upfront Design Thinking is best applied when there is a high level of uncertainty about the problem (i.e., stakeholder and user needs) and the corresponding solution. Creating Design Thinking-related artifacts through applying Design Thinking helps to understand the problem in depth and to define the overall concept of an idea. It is typically used at an early project stage to provide clarity for unclear user needs and to define a (high-level) solution vision (e.g., "How does the future patient support program for multiple sclerosis patients look like?").

*Prerequisites:* A problem statement should have been defined as a minimum starting point for applying upfront Design Thinking. Additional required conditions are the setup of a multidisciplinary project team, access to potential users and other stakeholders as well as Design Thinking training for project members.

*Key activities:* Design Thinking activities are typically performed in the form of a pre-project to identify relevant features that are worth implementing. The Design Thinking process model (define, needfinding, synthesis, ideation, prototyping, testing) guides through a cyclical creation of context and requirements artifacts. The outcome is used as a basis for performing further Requirements Engineering activities that complement Design Thinking artifacts with Requirements Engineering specific ones.

*Roles:* Two roles during the upfront mode are required. First, the Design Thinking team is responsible for planning and executing the activities. This team consists of four to six people from different areas of expertise depending on what knowledge will be relevant for the project, including for example subject matter experts, IT, marketing, sales, design personnel (Häger et al. 2015). Second, a person or group of people, who has defined the initial design challenge and project scope, is defined as the project sponsor. The person in this role typically provides continuous feedback to the team and connects it with others to enable synergistic effects and avoid duplicate efforts (Häger et al. 2015). The following two roles are optional: First, an extended team of (internal) experts that provide further domain knowledge and expertise for the Design Thinking team. Second, a Design Thinking coach or coaches who support the project team with methodological guidance. They introduce Design Thinking techniques, facilitate team meetings, and ensure that the team is focused on delivering the tasks and artifacts. As such, the coach should have a profound understanding of Design Thinking to provide useful techniques and guidance at appropriate times (Häger et al. 2015).

*Outcome:* The main deliverable of the upfront strategy is a clear system vision as a basis for performing further Requirements Engineering activities. The system vision usually takes the form of a mockup (i.e., high-fidelity prototype). Along the way the team will create a comprehensive set of Design Thinking artifacts, which should make it clear why each aspect of the prototype is intended in the way it is designed. High level user stories and a list of usability requirements based upon test results accompany the set of artifacts created by following the Design Thinking process.

---

*Case Example*

---

The international Alpha Insurance company wanted to develop a new service for their new target group of "young professionals". A project team stemming from five different business functions (marketing, IT, actuary, product manager, claims) spent 40% of their time to follow the Design Thinking process in an iterative manner for three months. The solution vision resulted in a tested medium-fidelity prototype for a digital on-demand insurance that could be activated and deactivated based on the user's preferences. The Design Thinking team handed over the prototype to the implementation team for further specification, testing, development, and market introduction. Transferred artifacts included a project documentation with twenty field studies, two personas, five opportunity areas, and six low-fidelity prototypes with learnings about failures. The final solution vision (in form of a mockup) specified key features and their usability. The implementation team performed tests to validate these features, their usability, and their service model.[4]

---

### 5.2.2 Infused Design Thinking

*Objective:* The main goal of this strategy is to support existing Requirements Engineering activities with selected Design Thinking techniques. This includes, for example, activities to clarify fuzzy requirements, foster creativity, gain new ideas, or to better understand user needs.

*Prerequisites:* The prerequisites for applying this strategy depend on the specific problem to be addressed. The problem should have a clear scope. The prerequisites as described in the previous still apply.

*Key activities:* An infused approach makes use of selected artifacts and leverages selected methods from the Design Thinking toolbox and integrates them into an existing Requirements Engineering process. In case of challenges encountered during the Requirements Engineering process, Design Thinking tools can be initiated; hence, their application is ad-hoc. The main activity of this strategy is the setup of focused workshops with selected Design Thinking tools (Dobrigkeit et al. 2108). These workshops can last three hours or several days depending on the objectives. For example, the goal of a workshop to generate new solution ideas could be formulated like this: "Create ideas to optimize the user interface of our platform, making it look and feel more emotional, and letting it appear less technical." This session used persona and customer journey artifacts to brainstorm new ideas.

---

[4] This case has also been published in Hehn et al. 2020 and Hehn 2020

*Roles:* In the infused setting, the people or person performing the Requirements Engineering activities are the addressees of receiving Design Thinking guidance in the form of workshops. Other workshop participants with different areas of expertise may be added, e.g., subject matter experts, IT, marketing, sales, design, depending on what knowledge will be relevant to achieve the workshop goal. A workshop typically consists of five to twenty participants. Like the upfront approach, a Design Thinking coach introduces the selected Design Thinking techniques and moderates the workshop and team discussions. The project sponsor can also be integrated to provide feedback and define the context for the general direction of the workshop.

*Outcome:* Due to the flexible approach of the infused strategy, the outcome is situation-dependent based on the previously defined objectives. The deliverables can be (new) features, user requirements, or test feedback – all following the Requirements Engineering process. In the context of the combined artifact model this means that the creation of Requirements Engineering artifacts is enhanced by a selected set of Design Thinking artifacts.

---

Case Example

---

Beta Enterprises is an international electronics group that wanted to evaluate the possibilities of smartphone applications (e.g., emergency apps, task lists, maintenance procedures) for container ships in a marine context. The main goal was to define requirements from a user point of view and to foster creativity for solution finding. In a highly regulated environment, a Design Thinking infusion was chosen to support the ongoing Requirements Engineering activities with selected tools from needfinding and prototyping. Five Design Thinking infusion sessions (one to two days) were conducted within five months. Produced artifacts included field studies for precise user requirements (it was the first time the team had been in close contact with marine captains) and tested medium-fidelity prototypes to strengthen service and usage models. According to the workshop participants, having direct user contact raised the confidence level in the success of the intended solution. Initial concerns about not finding interview partners in a highly sensitive B2B setting turned out as unjustified.[5]

---

### 5.2.3 Continuous Design Thinking

*Objective:* The main goal of this strategy is to integrate Design Thinking principles with Requirements Engineering activities on a continuous basis. Beyond the specific project context, this can also become part of an organizational change program or corporate strategy.

---

[5] This case has also been published in Hehn et al. 2020 and Hehn 2020

*Pre-requisites:* Continuous Design Thinking is recommended when addressing complex ("wicked") problem settings, which require continuous user involvement along all software engineering activities. In addition to the prerequisites described for the previous two strategies, (selected) project members should possess both Design Thinking and Requirements Engineering knowledge.

*Key activities:* Continuous Design Thinking utilizes the Design Thinking mindset as guiding principles. On an operational level, this translates into a seamless combination of the upfront and infused strategy and the potential setup of a new project role for a human-centric requirements engineer. The activities comply with both Design Thinking and Requirements Engineering elements to establish an end-to-end view from exploring a user's need to conceptualizing a solution vision and specifying a functional system. When starting a project, the upfront strategy can be used to provide clarity about the problem context and to elicit (user) requirements in a structured yet creative manner. A high-resolution prototype can help to specify the functionalities of the system vision. When moving on to the more technical side of requirements specification, an ad-hoc usage of Design Thinking methods can still be initiated in case features are not defined well enough from a user point of view for example.

*Roles:* The instantiation of a new role incorporates Design Thinking expertise as well as Requirements Engineering expertise and mediating between both schools of thoughts. In this strategy it is of great importance that the new role can react quickly when choosing methods and artifacts. The role enables the team to work towards a final product in incremental steps. The responsibilities of the project team during this strategy are like the preceding ones as the continuous strategy combines the two other strategies. The team plans and executes the activities to define the final system. The project sponsor has similar responsibilities as described in the previous sections.

*Outcome:* The continuous strategy results in a comprehensive set of Design Thinking and Requirements Engineering artifacts as shown in Figure 9. The requirements specification and system design are based on and traceable to customer needs derived from the context specification.

---

Case Example

---

Gamma Energy is a large energy provider with subsidiaries worldwide. A diverse project team applied an upfront Design Thinking approach to explore the potential of platforms in the utility sector. The outcome was a solution vision for a digital home improvement platform to advance lead generation. To ensure a human-centred mindset throughout specification and development, a new role was established to use selected Design Thinking tools for enhancing the prototype and filling the backlog with new features. Produced Design Thinking artifacts included high-fidelity prototypes with usability- and feature-oriented test feedback and new solution ideas. Scrum became the guiding framework for development, which enabled the entire project team to work in sprints. During development Design Thinking prototypes were used as boundary objects to enhance communication with relevant internal stakeholders and to foster a human-centred mindset within the team.[6]

---

## 5.3 Discussion

Our presented operationalization strategies reflect the ongoing discourse of describing Design Thinking at different levels in software engineering approaches (e.g., Brenner et al. 2016; Dobrigkeit and de Paula 2019). In line with other authors we suggest that the way in which Design Thinking should be used depends on the specific context and objectives of a project. Accordingly, three different strategies with different Design Thinking formats (e.g., process phases, workshops, single methods) were suggested which are similar to other proposed strategies in research in the context of (agile) software development. Depending on the situation each operationalization strategy offers different benefits but also challenges. Table 4 discusses both for each strategy.

---

[6] This case has also been published in Hehn et al. 2020 and Hehn 2020

**Table 4** Benefits and Challenges of each Operationalization Strategy (see also Hehn et al. 2020 and Hehn 2020)

| Strategy | Benefits | Challenges |
|---|---|---|
| Upfront Design Thinking | - The full potential of Design Thinking is leveraged while changes to Requirements Engineering are not necessary<br><br>- Due to the focus on problem exploration deep context understanding is achieved<br><br>- The solution concept has traceable links to user needs | - Resource- and time-intense<br><br>- Lost (implicit) knowledge and potential starvation of results when handing over Design Thinking results<br><br>- Little attention is paid to further development critical artifacts such as quality requirements, system constraints, or data models |
| Infused Design Thinking | - Intervention character requires only minimal changes in existing Requirements Engineering practices<br><br>- Resource and time friendly due to ad-hoc usage of selected tools (especially compared to upfront approach)<br><br>- Low adoption hurdle for Design Thinking methods | - Risk of neglecting problem understanding (especially compared to the upfront approach)<br><br>- No embedding of Design Thinking mindset due to situational Design Thinking workshops<br><br>- Little attention is paid to further development critical artifacts such as quality requirements, system constraints, or data models |
| Continuous Design Thinking | - Seamless integration into existing Requirements Engineering practices including development critical artifacts<br><br>- High likelihood of infusing a human-centred mindset within the project team<br><br>- Precise and traceable (user) requirements through continuous identification of new requirements and testing | - Requires commitment, resources, and time to develop continuous integration of both approaches in an organisation<br><br>- Continuous Design Thinking is highly dependent on the staffing of the project team<br><br>- Requires an organisational mind shift and support, potentially even an organisational restructuring |

Beside the project context, the existing maturity level of Design Thinking within an organization can be considered an influencing factor when choosing the 'right' strategy. While Requirements Engineering is usually an established practice in industry, Design Thinking is still relatively new. The decision to integrate the two approaches also depends on the level of courage, given time, and dedicated resources. As a rough guideline, the infusion strategy provides a reasonable starting point as it applies focused Design Thinking interventions within established practices. While the upfront strategy also keeps existing procedures, it requires more time and resources. Finally, the continuous strategy demands for a commitment from management to foster mindset change in an organisation or department.

A "morphing nature" of Design Thinking in software-intensive development projects can be stipulated, evolving from process-guidance, via toolbox support to the manifestation of a human-centered mindset of the project team. When approaching "wicked" problems, Design Thinking starts with a structured, upfront approach to define a clear product vision. Then, it transforms into a loose bundle of tools and a mindset that link well to common agile practices. Figure 13 visualizes this evolution.

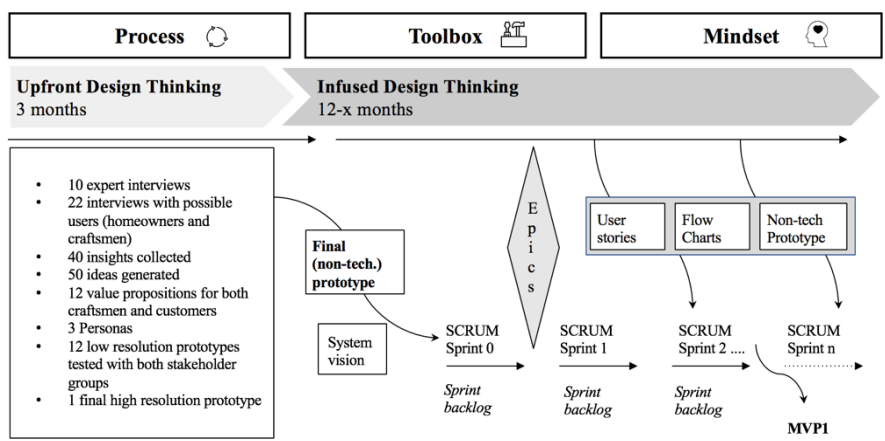

**Figure 13** Evolution from Process, via Toolbox, to Mindset (visualization based on findings from Hehn and Uebernickel 2018, see also Hehn 2020)

## 6 Synthesis of Findings

The following sections summarize our findings from sections 4 and 5.

### 6.1 Leveraging the Best of Both Worlds

Design Thinking and Requirements Engineering are not mutually exclusive but rather reinforce and complement each other. Using Design Thinking for Requirements Engineering means putting more focus on the early phases of the process to determine customer needs, requirements, and context, which affects the system vision with its product features and functionalities. Design Thinking expands the toolbox for Requirements Engineering by emphasizing artifacts for defining the relevance of the system vision. It fosters a holistic exploration of the problem context and defines precise user requirements. A prototype shapes the vision of the system. These artifacts complement the more technical-oriented artifacts from Requirements Engineering with a human-centered perspective. In addition, Requirements Engineering expands the toolbox of Design Thinking by connecting Design Thinking artifacts to later-staged software development processes. In this sense, Design Thinking-related artifacts are transformed into functionalities for technical realization. What counts in the end in Requirements Engineering is the set of elaborated requirements, while in Design Thinking, not only the prototype is the ultimate outcome, but the learning curve leading to it.

For creating a lasting impact of the system vision on the upcoming design and implementation activities, a balance should be found between the benefits of early experimentation as done in Design Thinking and the advantages of institutionalizing a proper structure and documentation for subsequent software engineering activities as achieved by Requirements Engineering.

## 6.2 A Comprehensive Blueprint for Innovative Software-intensive Systems

We contribute an evaluated artifact model for Design Thinking and Requirements Engineering that can be tailored to specific project situations. The model is descriptive and prescriptive at the same time. It is descriptive by depicting the most common Design Thinking and Requirements Engineering artifacts as used in software-intensive development projects. It can be seen as a blueprint for designing new innovative systems, which makes it also prescriptive as it provides a guideline and orientation for generating the artifacts in development projects. Managers can use the model to evaluate their Requirements Engineering processes and, thereby, improve effectiveness and create solutions in a more human-centred fashion.

## 6.3 There is no "One Size Fits All"-Integration Strategy

Operation modes that integrate Design Thinking into (agile) software development approaches have been proposed before (e.g., Lindberg et al. 2012; Häger et al. 2015; Dobrigkeit et al. 2018). Building on these findings and triangulating them with empirical data from industry, three operationalization strategies have been identified in which Requirements Engineering can profit from Design Thinking and vice versa: (1) Run Design Thinking upfront to Requirements Engineering practices (upfront Design Thinking), (2) infuse Requirements Engineering with selected Design Thinking tools (infused Design Thinking), or (3) apply Design Thinking and Requirements Engineering continuously in a flexible manner (continuous Design Thinking). The decision which strategy to follow depends on the project context and objective. The first strategy is recommended when the problem and solution space is unclear. Here, the Design Thinking process provides a guiding structure for requirements elicitation and the specification of a solution vision. The second strategy offers requirements engineers a way to make use of selected Design Thinking methods when they feel it is necessary. Typically, these are situations in which project members face difficulties in an ongoing Requirements Engineering process that might be addressed by Design Thinking methods. The third strategy supports a continuous yet flexible application of the Design Thinking process and ad-hoc tools. The continuous approach entails the evolution from using Design Thinking as a guiding process to applying it as a toolbox for adaptive support up to implementing Design Thinking principles in the mindset of project members. This strategy should be chosen when (1) a sustainable integration of both Design Thinking and Requirements Engineering is intended and (2) the project requires a continuous integration of users into the development project. In this context, the human-centric requirements engineer is a new role that incorporates skills from both disciplines. Business analysts may leverage Design Thinking to deeply explore the system context while design thinkers may equip themselves with Requirements Engineering knowledge to better connect their results to subsequent software design.

## 6 Conclusion

Design Thinking offers great potential for promoting innovative, user-centered concepts as it promises to place users and their needs at the core of the design process. This gave rise to great interest in using Design Thinking for the engineering of software-intensive systems and services which are nowadays challenged by their pervasive nature, ever-growing complexity, and the inherent difficulty to capture requirements and development constraints in a user-centric manner. Despite the popularity of Design Thinking in research and practice, it is, however, often treated in isolation without much care for a clear, seamless integration into established software engineering approaches. In fact, too often, we tend to pretend that problem solving ends with a deeper understanding of the problem domain and by building mostly non-technical prototypes and, thus, leaving open an effective transition into actual development and quality assurance. At the same time, in software engineering research and practice, we pretend too often that requirements are just there and that they simply need to be elicited and documented (if at all) and, thus, missing out great potential of fully exploring the problem space in a human-centric manner.

The idea of integrating Design Thinking into Software Requirements Engineering approaches to leverage the potential of a deeper problem exploration and discovering and specifying requirements more thoroughly is not new. However, Requirements Engineering and Design Thinking come both in various forms and interpretations rendering such an integration cumbersome. Thus, integration efforts typically end at the high level of abstract principles, values and mindsets, and practices. In this chapter, we therefore took an artifact-centric perspective to (1) synthesize both at a terminological and conceptual level, and to (2) lay the foundation of effectively guiding the problem-oriented specification of requirements based on a seamless and holistic underlying artifact model. Our contributions focus on the following two aspects:

- We elaborated on the very fundamental principles and practices of both Design Thinking and Requirements Engineering and established two independent artifact models that reflect those principles. Here, we drew from both the state of the art in Design Thinking and in Requirements Engineering as well as from experiences made along two decades of academic-industry collaborations.
- We integrated both, the artifact model for Design Thinking and the model for Requirements Engineering and presented different operationalization strategies of how to make efficient use of that integrated approach to create human-centered software-intensive systems.

Note that rather than merely focusing on a purely academically oriented model, we aimed at elaborating on essential terms, principles, and concepts while considering and extending the perspective on the practical relevance as many results emerge from academia-industry collaborations. The choice of the artifact-centric, process-agnostic philosophy further served two major purposes. First, it allowed us to lay such conceptual and terminological foundation for an integrated approach while, second, not enforcing a rigid, pre-defined structure for one (and only one) specific way of working (and thinking), hence, accommodating the various project situations and disciplinary backgrounds we face.

This is in tune with the overall scope of this book. We aim at creating a space to further foster debates and efforts in integrating both Design Thinking and Software and Systems Engineering by inviting scholars and practitioners from both interdisciplinary communities while not enforcing respective historically grown worldviews on each other. One hope we associate with this introductory chapter as well as with the overall book is to motivate the value of such an integration of both worlds.

**Acknowledgements**

We would like to thank Falk Uebernickel for his continuous support and feedback in previous articles and research efforts that provided major influence on our findings presented in this book chapter. We further thank Manfred Broy and Walter Brenner for stimulating discussions and feedback on earlier versions of this manuscript.

# Appendix

## A. Artifact Description

The following appendix defines the content model of the combined artifact model in detail giving for each content item a definition of the used concepts.

The *Number (#)* references the assigned number within the artifact model.

The *Name* captures the name and the type of the artifact. If the artifact can be attributed to both Design Thinking (DT) and Requirements Engineering (RE), different descriptions for both approaches (e.g., Design Challenge and Project Scope) are marked by a slash (/). In this case, the description for the Design Thinking-related artifact is provided first and the Requirements Engineering expression second.

*Description & Purpose* denotes the content and main characteristics of each artifact type. Interdependencies summarize the relationships between the artifacts regarding their content within the artifact model. The description differentiates between the input that artifacts receive from the content of other artifacts ('input from') and the output that they provide for other artifacts in the artifact model ('input for').

The *Notation* suggests appropriate documentation and specification techniques for each artifact (e.g., natural language, Unified Modelling Language (UML) class diagrams, model-based documentation).

### A.1 Context Specification

A description of the content items of the context specification is provided in Table 5.

**Table 5** Content Items in the Context Specification

| # | Name | Description & Purpose | Notation |
|---|------|----------------------|----------|
| 01 | Design Challenge / Project Scope (DT&RE) | Describes the business problem and provides direction for problem analysis and development; has an exploratory character in DT, a convergent objective in RE. Input for (#05), (#08), (#27), (#30). | Natural Text |
| 02 | Constraints and Rules (DT&RE) | Restrictions and fixed design decisions that influence the system design and implementation and must be obeyed or satisfied; establishing them helps to run and manage the project within the intended business and technical restrictions; constraints are often explicitly challenged in DT. Input for (#05) | Natural Text |

| 03 | Business Case (DT&RE) | Provides rationale for a design project and is used to convince decision-maker or project sponsor; in DT its main objective is to evaluate available execution budget (resources and time), in RE it may have concrete solution options in mind.<br>Input from (#01); input for (#05) | natural text |
|----|-----------------------|----------------------------------------------------------------------------------------------------|--------------|
| 04 | Stakeholder Map / Stakeholder Model (DT&RE) | List of relevant stakeholders (internal and external) for the project, typically including project sponsor or client, project manager, product manager, other (senior) decision-makers, investors, end users, customers, operators, product disposers, sales and marketing, or regulatory authorities; helps to identify key internal and external stakeholders as sources of requirements.<br>Input from (#01); input for (#05), (#07), (#25) | natural text, diagram, UML actor hierarchy |
| 05 | Objectives and Goals (DT&RE) | Prescriptive statements of intent regarding business, usage, or system goals issued by a stakeholder (e.g., quality-related, optimization-specific, behavioural, anti-goals); provide direction for problem analysis and system development; in DT the list contains mainly high-level business goals and objectives provided by the project sponsor to keep outcome and specifics open for exploration; in RE they may be more precise.<br>Input from (#01), (#02), (#03), (#04); input for (#06), (#24), (#25) | natural text, goal graphs |
| 06 | Domain Model (RE) | Composed of all real-life conceptual objects related to a specific problem (incl. business entities, attributes, roles, relationships, constraints); ensures an understanding of the landscape of business entities in the problem area and can be used to solve problems related to that domain.<br>Input from (03#), (05#); input for (#09, #34, #40), (#24), (#25) | UML activity diagrams; Business Process Model Notation (BPMN) |
| 07 | Design Space Map (DT) | Overview of knowledge and knowledge gaps in the context of the project; helps to structure the exploration phase and provides a common understanding of the design challenge; it evolves over the duration of a project in which new knowledge is added.<br>Input from (#01), (#04); input for (#10), (#11) | natural text |

| 08 | Assumptions (DT) | Hypotheses about project and stakeholders to be explored and tested in the project; provides a first overview of possible team biases.<br>Input from (#01), (#04); input for (#17), (#18) | natural text |
|----|----|----|----|
| 09 | Glossary (RE) | List of all relevant business or technical domain-specific terms to ensure their consistent usage throughout the entire development life cycle; key elements are terms, definitions, aliases, and related terms<br>Input from (#04), (#06), (#07); input for (#34), (#40) | natural text |
| 10 | Secondary Research Report (DT) | Summary of various sources of information and insights from existing market research about the given subject domain (e.g., market and benchmarking reports, sales reports, internal databases, government statistics, articles, research studies); the report supports the project team to clarify research questions and gain an initial understanding of the challenge context<br>Input from (#07); input for (#12), (#15) | natural text |
| 11 | Field Studies (DT) | Collection of raw data (incl. statements, observations, pictures, videos) from interviewees; they help the team to create a common understanding of the raw data and empathize with the interviewees.<br>Input from (#04), (#07); input for (#12), (#13), (#14) | natural text, pictures, videos |
| 12 | Thematic Clusters (DT) | Group of user statements, observations, and other findings from primary and secondary research that represent a specific subtopic of the project content; they provide an overview of relevant topics within a given domain and help the project team to recognize patterns<br>Input from (#11), (#10); input for (#15). | natural text |
| 13 | Personas (DT) | Fictional characters that represent a specific stakeholder group relevant to the project (incl. a demographic profile, behavioural patterns, attitudes, goals); they facilitate the understanding of (potential) users' needs, behaviours, motivations, and frustrations and provide alignment for discussing design decisions.<br>Input from (#11); input for (#14), (#16), (#17) | natural text; pictures |

| 14 | Customer Journeys (DT) | Visual representations of the experience of a customer when interacting with an organisation, product, or service (activities, tasks, touchpoints); they offer a systematic analysis of challenges, pain and gain points that help to identify areas with innovation potential<br>Input from (#11), (#13); input for (#15), (#16), (#25) | natural text; sequence & activity diagrams |
|---|---|---|---|
| 15 | Insights (DT) | Findings that occur because of synthesis and interpretation of primary research; usually expressed in one sentence to explain why something is happening<br>Input from (#11), (#12); input for (#16), (#17). | natural text |
| 16 | Opportunity Areas (DT) | Potential for innovation based on insights and needs found in primary research; they define specific directions for next steps while they often go beyond the project assignment itself; the formulation of opportunity areas is rather action-oriented, while the insights describe the status quo or a desired future state.<br>Input from (#12) - (#15); input for (#17) | natural text |
| 17 | Solution Ideas (DT) | Specific features and concepts on how to solve a given problem statement (based on creativity techniques and brainstorming)<br>Input from (#09); input for (#18), (#20), (#22) | Natural text |
| 18 | Low-fidelity Prototypes (DT) | Tangible and testable artifacts that demonstrate the key functionalities of an idea; examples are paper prototypes, role plays, Wizard of Oz; particularly suitable during the early stages of a project, when the topic is still abstract or in the process of forming as costs and effort are extremely low, which allows the project team to explore various ideas at once<br>Input from (#17); input for (#18), (#20) | different forms, mostly in a paper-based format |
| 19 | Scope-oriented Test Results (DT) | Feedback from users and other relevant stakeholders regarding the basic concept of an idea; it helps the team to gain more empathy for their target group and to decide which ideas to keep, to refine, and to drop<br>Input from (#18); input for (#20), (#22) | natural text |
| 20 | Medium-fidelity Prototypes (DT) | Non-technical prototype showing key features of the target product or service; while low-fidelity prototypes (#18) are useful to inspire new ideas, medium-fidelity prototypes are mainly used to test and refine existing solution ideas; they usually take more effort to build, yet also provide a much more realistic representation of the envisioned behaviour and user interface.<br>Input from (#17), (#18), (#19); input for (#21), (#22) | different forms, mostly in a digital format |

| 21 | Feature-oriented Test Results (DT) | Feedback from users and other relevant stakeholders regarding key features and functionalities of the prototype; they validate customer's expectations and help to prioritize functionalities for implementation | natural text |
|----|----|----|----|
| | | Input from (#20); input for (#22), (#25, 26) | |

## A.2 Requirements Specification

A description of the content items of the requirements specification is provided in Table 6.

**Table 6** Content Items in the Requirements Specification

| # | Name | Description & Purpose | Notation |
|---|------|---------------------|----------|
| 22 | High-fidelity Prototypes (DT) | Offers a clear vision of how the final system will look and feel; they help the project team to gain meaningful feedback for usability testing and are also suitable to gain buy-in from clients and internal project stakeholders<br>Input from (#17), (#18) - (#21); input for (#22), (#24). | different forms, mostly in a digital format |
| 23 | Usability-oriented Test Results (DT) | Feedback from users and other relevant stakeholders regarding the interaction with a product; the results provide areas for improving issues of understandability and point at directions for refining design elements and interaction mechanisms<br>Input from (#22); input for (#24), (#25) | Natural Text; pictures, videos |
| 24 | System Vision (DT&RE) | Specification of how an information system is to fit into the business context while supporting pre-defined restrictions and goals; it serves as a means for agreeing on what the solution is about; while the purpose of the system vision is similar to both DT and RE, its realization might be different: in DT it is usually comprised of a high-level natural text specification and a medium-or high-fidelity prototype (#20, #22), in RE the system vision is often expressed via rich picture.<br>Input from (#03), (#04), (#05), (#06), (#22); input for (#25), (#33), (#31) | rich picture, prototype, natural text |
| 25 | Usage Model (DT&RE) | Illustration of the (black box) system behaviour of the system vision (#24) from the user's point of view through an overview of use cases (incl. actor, task, objective, and causal relationship); the model provides an understanding about which system functions are performed for which actors (in their roles); while the purpose of the usage model is similar to both DT and RE, its realization might be different.<br>Input from (#13), (#14), (#24), (#26); input for (#28), (#29), (#33). | natural text, UML activity diagrams |

| 26 | Service Model (DT&RE) | Specification of requirements and objectives of the intended services of the solution (i.e., user-visible functions through input/output-relations); it provides a comprehensive understanding of the services and their underlying resources and processes, whether seen or unseen by the user; while the purpose of the usage model is like both DT and RE, its realization might be different. Input from (#24); input for (#25), (#29), (#33) | natural text; graphs |
|---|---|---|---|
| 27 | Process Requirements (RE) | Activities that should be performed by the developing organisation (e.g., compliance to standards and process models, project milestones, style-guides, infrastructure); they provide the guidelines for a consistent design and implementation of the intended system Input from (#01) | natural text |
| 28 | Functional Hierarchy (RE) | Specification of functions and subfunctions and their relationships and dependencies; functions are user-visible pieces of the system behaviour that correspond to services in (#26) and realize system actions from (#25); bridges the requirements and system specification and can be used as a guideline for obtaining and organizing system requirements Input for (#29), (#36), (#38), (#39) | graphs & input-output tables |
| 29 | Data Model (RE) | Summary of all data objects and relations that are part of the system's functions and interaction scenarios; it supports the development of the intended system by providing the definition, format, and structure of the required data Input from (#25), (#26), (#28); input for (#37) | UML class diagrams |
| 30 | Deployment Requirements (RE) | Description of demands for making the software available for use, i.e. specifying the process of the deployment and the technical infrastructure during the initial release of the system or specific parts of it; they contribute to the overall quality of the resulting system Input from (#01) | natural text |
| 31 | Risk List (RE) | Description of all risks that are related to project-specific requirements and that potentially threaten the development or operation of a system; risks are typically analyzed along stakeholder interests and estimated regarding their probability and potential damage; the risk list provides the foundation to introduce necessary countermeasures | natural text |

| | | Input from (#24) | |
|---|---|---|---|
| 32 | System Constraints (RE) | Logical and technical restrictions for the system architecture, its functionality, and quality; they provide the boundaries for development and deployment<br>Input for (#38). | natural text |
| 33 | Quality Requirements (RE) | Desired quality characteristics of a system beyond functionality and features (e.g., reliability, performance, security, usability, adaptability); they are assessed by pre-defined measurements and help to validate the successful completion of an entire system or its respective functions and features<br>Input from (#11), (#13), (05#), (#24), (#25); input for (#36), (#38). | natural text |
| 34 | Glossary (RE) | Extends the glossary of context-relevant terms (#09) with requirements-specific terms; it will show up again in the system specification (#40) as more terms are added | natural text |
| 35 | Architecture Overview (RE) | Aggregation of component overview (#38) and functional hierarchy (#28); offers high-level understanding of the evolving system's architecture and guides the definition of the more intricate functional and operational architecture.<br>Input for (#36), (#38). | component diagram |

### A.3 System Specification

A description of the content items of the system specification is provided in Table 7.

**Table 7** Content Items in the System Specification

| # | Name | Description & Purpose | Notation |
|---|------|----------------------|----------|
| 36 | Function Model (RE) | Overview diagram of the user-observable functions and their communication relationships; the model ensures an overview of all functions and processes and, thus, assists in determining the scope for implementation and the product and service costs Input from (#28), (#33), (#35), (#38); input for (#39) | graphs, tables |
| 37 | Data Model (RE) | Overview of the coarse-grained data objects and the relations that are required for the executing the system's functions; the "data elements" of the data model refine the "data objects" from the data model (#29) in the requirements layer by using a particular data type; it is part of a stepwise completion from moving the focus on defining user-visible functions towards specifying the design system Input from (#29); input for (#39) | UML class diagrams |
| 38 | Component Model (RE) | Description of the components (i.e., building blocks) of a system's services and their respective channels and interfaces (e.g., application components, system software components, technical components, hardware components); the model bridges the requirements layer with the system layer by defining the main design principles and overall structure of the system Input from (#32), (#33), (#35); input for (#36), (#39) | component diagrams |
| 39 | Behaviour Model (RE) | Description of the internal behaviour of a system with the goal to execute the defined functionalities; the model depicts a dynamic view of the system behaviour and illustrates how objects or system components interact to support use cases Input from (#25), (#36), (#38); input for (#37) | Inter-action diagrams, behavioural state machines |
| 40 | Glossary (RE) | Extends the previously defined glossary artifacts (#09, #34) with technical relevant terms. | natural text |